\documentclass[showpacs,twocolumn]{revtex4-1}
\usepackage{graphicx}
\usepackage[]{epsfig}

\newcommand{\beq}{\begin{equation}}
\newcommand{\eeq}{\end{equation}}
\newcommand{\beqd}{\begin{displaymath}}
\newcommand{\eeqd}{\end{displaymath}}
\newcommand{\beqa}{\begin{eqnarray}}
\newcommand{\eeqa}{\end{eqnarray}}

\newcommand{\comment}[1]{}

\begin{document}
\title{On The Nature of the Glass Crossover}
\author{Tommaso Rizzo}
\affiliation{Dip.\ Fisica, Universit\`a ``Sapienza'', Piazzale A.~Moro 2, I--00185, Rome, Italy}
\affiliation{ISC-CNR, UOS Rome, Universit\`a ``Sapienza'', Piazzale A.~Moro 2, I-00185, Rome, Italy}
\author{Thomas Voigtmann}
\affiliation{Deutsches Zentrum f\"ur Luft- und Raumfahrt (DLR), 51170 K\"oln, Germany}
\affiliation{Department of Physics, Heinrich-Heine-Universit\"at D\"usseldorf, Universit\"atsstra\ss{}e 1, 40225 D\"usseldorf, Germany}


\begin{abstract}
Stochastic Beta Relaxation (SBR) is a model for the  dynamics of glass-
forming liquids close to the glass transition singularity of the idealized mode-
coupling theory (MCT) that has been derived from generic MCT-like theories by applying dynamical field-theory techniques. SBR displays a rich phenomenology common to most super-cooled liquids. In its
simplest version it naturally explains two prominent features of the dynamical
crossover: the change from a power-law to exponential increase in the
structural relaxation time and the violation of the Stokes-Einstein relation
between diffusion and viscosity.
The solution of the model in three dimensions unveils a qualitative change at the crossover in
the structure of dynamical fluctuations  from a regime characterized by power-law increases of their amplitude and size to a regime dominated by strong Dynamical Heterogeneities: rare regions where dynamics is relatively much faster than in the rest of the system.  
While the relaxation time  changes by orders of magnitude, the size of these regions does not change significantly and actually decreases below the crossover temperature.
SBR cannot sustain too large fluctuations and could fail well below the crossover temperature. There it could be replaced by  non-conventional activated dynamics characterized by elementary events with intrinsic time and length scales of an unusual large (but not necessarily increasing) size (mesoscopic vs. microscopic). 
\end{abstract}

\maketitle

\section{Introduction}

Mode-coupling-Theory provides a good qualitative and quantitative description of the initial dynamical slowing down of super-cooled liquids but its main problem is that it predicts dynamical arrest at a temperature $T_c$ where a crossover is observed instead \cite{Goetze09}.
Recently it has been shown that the solution to this problem may come by treating the singularity as a genuine phase transition by means of perturbative field-theoretical methods \cite{Rizzo14}. This is surprising because in general perturbative methods are unable to remove a singularity, however one can show that in the case of MCT the perturbative loop corrections are the same of those of some dynamical stochastic equations called stochastic-beta-relaxation (SBR) equations in \cite{Rizzo14}. If studied perturbatively both MCT-like theories  and SBR display dynamical arrest at all orders, but the SBR equations can be also solved explicitly ({\it i.e.} non perturbatively) showing that the transition is instead changed into a crossover due to non-perturbative effects that can be clearly identified.

The result is rather intuitive: on the time-scale of the $\beta$-regime (where by definition the density-density correlator remains close to a plateau) dynamics according to the SBR equations is described by the very same MCT critical equations, the only difference being that the temperature fluctuates randomly between different regions of the system. As a consequence, even if the global temperature is near $T_c$ or below, there are regions of the system in the liquid phase and they destabilize the glass phase predicted by ideal MCT and restore ergodicity.

In a previous publication \cite{Rizzo14b} we have studied a sort of schematic version of the model where
different regions  are uncorrelated and behave independently. The study of the model requires elementary computations but, notwithstanding its simplicity, displays many features typically observed in super-cooled liquids and allows to understand them in an intuitive way. In particular when supplemented with the assumption of time-temperature superposition this simplified SBR allows to obtain predictions on the $\alpha$ regime that are in remarkable agreement with the known phenomenology of various quantities, including the $\alpha$-relaxation time, the Diffusion constant and the thermal susceptibility.
The main limit of the simplified model is that it does not allow to study length scales which requires the solution of the full SBR equation in finite dimension.  

In this paper we will discuss the solution of the SBR equations in three dimensions.
From the solution one can draw a rather comprehensive description of the glass crossover that we will sketch in the following. At any point in space we associate some local field that quantifies somehow the mobility {\it i. e. } the {\it time rate} with which that portion of the system decorrelates from the initial configuration.
The amplitude and the size of the fluctuations of the local mobility allows to quantify and discuss the notion of Dynamical Heterogeneities.  The solution of the SBR equations suggest that 

\begin{itemize}
\item approaching $T_c$ from above the mobility field (associated to the function $B(x)$ in the following) decreases in average value (it would be zero in ideal MCT at $T_c$) while its local fluctuations increases both in size and in amplitude.
More precisely the process has the features of a second-order scale-invariant phase transition and in particular it is characterized by an increasing dynamical correlation length. However the increase of the relative fluctuations is not very pronounced approaching $T_c$ from above and it seems not appropriate to talk of Dynamical Heterogeneities in the sense used by experimentalist.

\item Close to $T_c$, there is a change and two important things happens: i) dynamical arrest is avoided and ii) the relaxation time  starts to grow much faster, from power-law to exponential-like.

\item The structure of dynamical fluctuations also displays a qualitative change below $T_c$: overall the dynamics continue to slow down dramatically but it is dominated by rare regions that are relatively much faster than the rest of the system. Dynamical Heterogeneities are the hallmark of the dynamics in this regime. 

\item Below $T_c$ dynamics is slow not because the faster regions are large, but rather because they are rare and therefore the amplitude of fluctuations increases.
Actually the size of these regions shrinks while decreasing the temperature below $T_c$ and the dynamical correlation length decreases. 
In general the change in the correlation length is not dramatic and decouples from the increase of the relaxation time. The key point is that the structure of the fluctuations changes from scale-invariant-like above $T_c$ to activated-like below $T_c$. Correspondingly quantities that scales similarly above $T_c$ decouples leading {\it e.g.} to deviations from  the Stokes-Einstein relationship (SER).

\end{itemize}

In the next section we will discuss how the above picture emerges from the numerical solution of the SBR equations in 3D and in the final section we will discuss the results.

\section{A Theory of the Glass Crossover}

\subsection{Stochastic Beta Relaxation}

MCT and similar mean-field theories lead to the prediction that near the critical temperature the density-density correlator has the following behavior on the time scale of the $\beta$-regime  $\tau_\beta$:
\beq
{\bf \Phi}(k,t)={\bf F}(k)+G(t) \, {\bf H}(k)\, 
\label{scalfor}
\eeq
where the bold character accounts for the case of mixtures of particles in which the correlator is a matrix.
The function $G(t)$ obeys the well-known MCT equation for the critical correlator:
\begin{displaymath}
\sigma=-\lambda \, G^2(t) +{d \over dt}\int_0^t G(t-s)G(s)ds
\end{displaymath}
where the separation parameter $\sigma$ is negative at high temperatures (low pressures) and vanishes at the MCT singularity: 
\begin{center}
$\sigma \propto T_c-T,\, \sigma \propto \rho-\rho_c$ 
\end{center}
The solution of the above equation is such that $G(t)$ goes to minus infinity at large times in the liquid phase according to the so-called Von Schweidler's law:
\begin{displaymath}
\lim_{t \rightarrow \infty} G(t) = -B(\sigma) t^b \ \ \mathrm{for} \  \sigma <0 \   \mathrm{(liquid)}
\end{displaymath}
and it goes instead to a constant in the glassy phase, signalling that the correlator remains blocked near the ergodicity breaking parameter ${\bf F^c}(k)$:
\begin{displaymath}
\lim_{t \rightarrow \infty}G(t) =\sqrt{\sigma/(1-\lambda)} \ \ \mathrm{for} \  \sigma >0 \   \mathrm{(glass)}
\end{displaymath}
The exponent $b$ is expressed in term of the exponent parameter $\lambda$ through: 
\beq
\lambda={\Gamma^2(1-a)\over\Gamma(1-2a)}={\Gamma^2(1+b)\over\Gamma(1+2b)}
\label{lambdaMCT}
\eeq
Where $a$ is the exponent controlling the small time behavior of $G(t) \sim 1/t^a$both in the liquid and glassy phase.
The critical equation is valid provided that $G(t)$ is small and this condition defines the $\beta$ regime. When $G(t)$ becomes $O(1)$ we enter the $\alpha$ regime, whose time-scale can therefore be obtained as:
\beq
\tau_{\alpha}  \propto B(\sigma)^{-1/b}
\eeq 
The prefactor $B(\sigma)$ of the $-t^b$ term vanishes in ideal MCT approaching $\sigma=0$ as $|\sigma|^{\gamma b}$ (where $\gamma=1/(2a)+1/(2b)$) and as a consequence $\tau_\alpha$ diverges as $\tau_\alpha \sim  |\sigma|^{-\gamma}$  . 

SBR can be viewed as an extension of the MCT equation for
the critical correlator with random fluctuations of the separation parameter.
According to it in the $\beta$ regime equation (\ref{scalfor}) continues to hold and only the epxression of the critical correlator $G(t)$ is different.
One must consider a field $g(x,t)$ that is a local version of the correlator and that obeys the following equation:
\beq
\sigma  + s(x) =- \alpha \nabla^2 \, g(x,t)-\lambda \, g^2(x,t)+{d \over dt}\int_0^t g(x,t-s)g(x,s)ds
\eeq
where the field $s(x)$ is a {\it time-independent} random fluctuation of the separation parameter, Gaussian and delta-correlated in space:  
\beq
[s(x)]=0\, ,\ [s(x)s(y)]=\Delta \sigma^2  \, \delta (x-y) 
\eeq
The  total correlator is obtained as the integral over space averaged over the random fluctuations:
\vspace{10px}
\beq
G(t) = [{1 \over V}\int g(x,t) dx]
\eeq
Thus SBR introduces two novel parameters in the description of the $\beta$ regime: the variance $\Delta \sigma^2$ of the random fluctuations and the length-scale of spatial correlation which are controlled by the coefficient $\alpha$ in front of the nabla term.
The fact that SBR can be derived from MCT-like theories at criticality implies that quantitatively these parameters can be computed within MCT similarly to $\lambda$ and $\sigma$ \cite{Rizzo14}, for instance the parameter $\alpha$ can be extracted from Inhomogeneous MCT \cite{Biroli06}. Different estimates can be obtained in native field-theoretical approaches \cite{Franz12}.

\subsection{The $B$-profile}

In the simplified model the dynamics in each region depends solely on the  given value of the local temperature and this allows to understand two important features of SBR \cite{Rizzo13,Rizzo14,Rizzo14b}. First of all one can see that at any temperature, even deeply below $T_c$ there will be liquid regions because of fluctuations of $s(x)$, therefore dynamical arrest is avoided. Second, the typical region is liquid above $T_c$ but is frozen below $T_c$ and correspondingly the fluctuations $s(x) \sim -\sigma$ required to have a liquid region become increasingly rare below $T_c$; this determine a crossover in the growth of the relaxation time from power-law to exponential.
\begin{figure}
\begin{center}
\includegraphics[width=8.5cm]{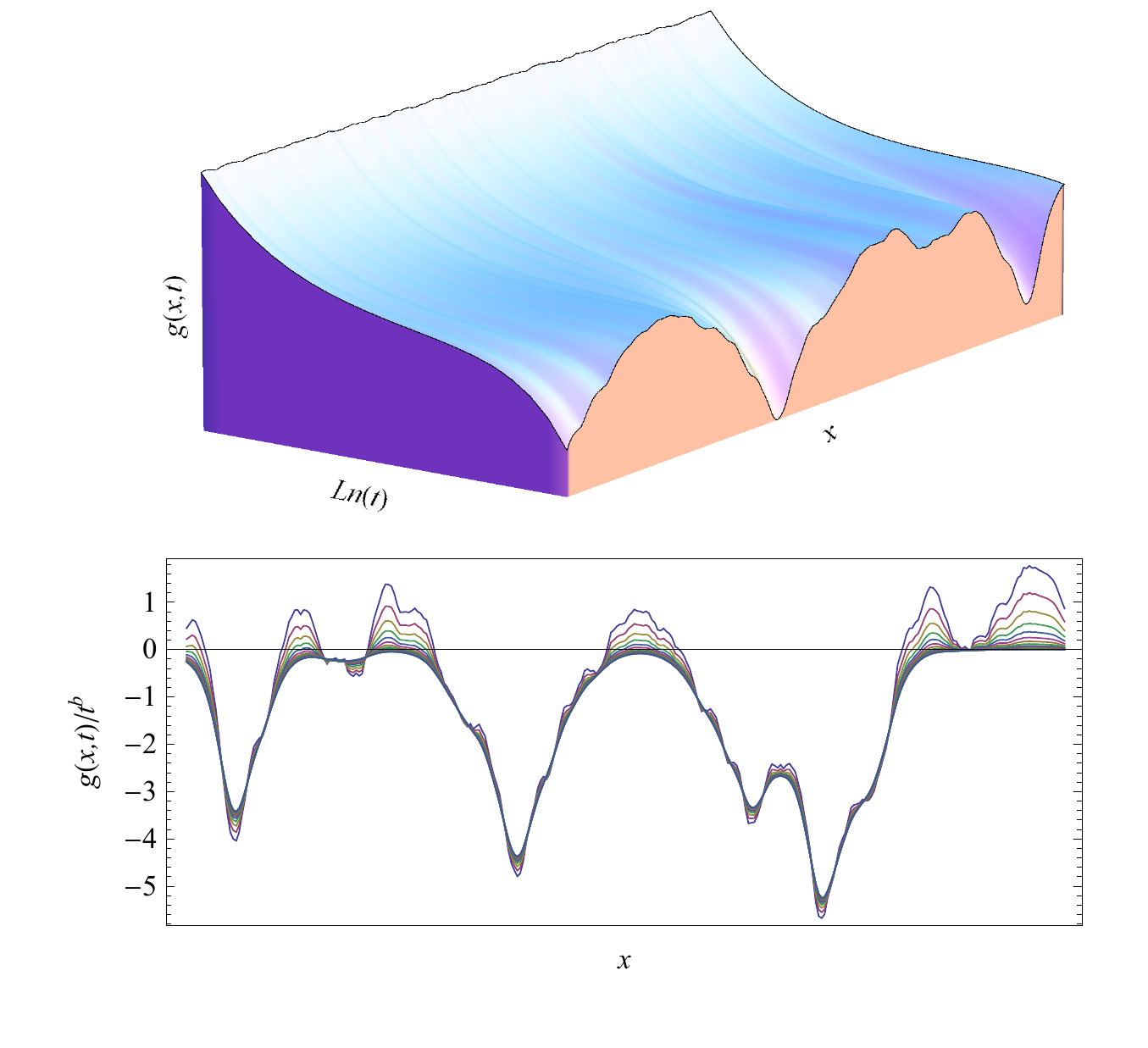}
\caption{Top: Pictorial representation of the solution of SBR equations in finite dimension for a given realization of the random $s(x)$.  (generated from actual solution of a 1D system). At large times the solution converges to the form $-B(x)t^b$, thus inducing a mapping between the realization of the random $s(x)$ and a positive function $B(x)$, the $B$-profile. Bottom: Plot of $g(x,t)/t^b$ vs. $x$ for increasing times $t$: increasing the time the curves converge to  the $B$-profile.}
\label{fig:plot-B.pdf}
\end{center}\end{figure}
We have studied the SBR equation in 3D numerically solving for $g(x,t)$ at given $s(x)$ and verified that the above description remains essentially valid.
In particular both for negative and positive values of $\sigma$ the solution $g(x,t)$ escapes to minus infinity ({\it i.e.} leaves the plateau) at all points $x$.
This corrects a disturbing pathology of the simplified SBR equations where there are always regions that remains blocked near the plateau and never decay.
More precisely $g(x,t)$ exits from the plateau with the very same Von-Schweidler law
with a space dependent constant $B(x)$, (see fig. \ref{fig:plot-B.pdf} )
\begin{displaymath}
g(x,t) \sim  -B_{\sigma}(x) t^b \ \ \ for\ all\  \sigma  
\end{displaymath}
Therefore at any value of $\sigma$ the SBR equations induce a non-trivial mapping between the realization of the fluctuations $s(x)$ and a positive function $B(x)$ (called $B$-profile in the following). 
In practice we extract the $B$-profile by solving the equations up to times large enough to be in the asymptotic regime where $g(x,t) \approx -B(x)t^b$.
The presence of the gradient term in the full SBR equation leads to the disappearance of a clear distinction between  liquid and glassy regions as all regions become liquid,  nevertheless when we switch on the gradient the regions that were glassy will be characterized by a very small value of $B(x)$ compared to the rare liquid regions where $B(x)$ is relatively much larger.
The $B$-profile will be our main focus in the following, indeed it allows a discussion in a clear and compact way of many dynamical quantities including the $\alpha$-relaxation time, diffusion coefficient, correlation length and Dynamical Heterogeneities.

It is illuminating to directly inspect the $B$-profile above and below the critical temperature. 
\begin{figure}
\begin{center}
\includegraphics[width=8.5cm]{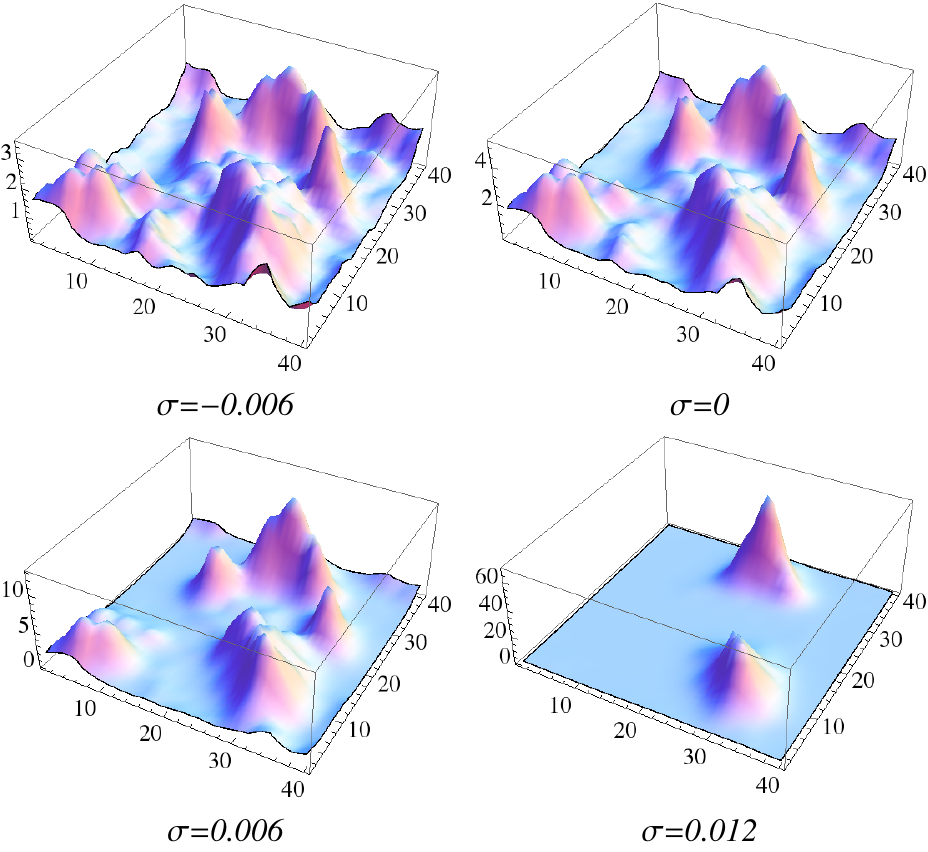}
\caption{Normalized $B$-profile $B(x)/\overline{B}$ on a plane sliced from a cubic box for different values of $\sigma$.}
\label{fig:B-slices}
\end{center}\end{figure}
In figure (\ref{fig:B-slices}) we plot the normalized $B$-profile $B(x)/\overline{B}$ on a plane sliced from a cubic box. The peaks(valleys) in the $B$-profile correspond to regions that are decorrelating from the initial condition faster(slower) than the average and  where local dynamics is also faster(slower). The height and size of the peaks allows to characterize dynamical fluctuations. We see that for  $T \geq T_c$ ( {\it i.e.} $\sigma \leq 0$) fluctuations are not very pronounced and the system appear homogeneous, instead below $T_c$ more and more peaks disappear and as a consequence the few that are left tend to be much more pronounced. Overall the average value $\overline{B}$ decrease monotonically  with decreasing the temperature (it changes by orders of magnitude for the values of $\sigma$ of the plot) but the dynamics becomes instead dominated by rare regions that are relatively much faster than the typical region. Thus the evolution of the $B$-profile encodes a qualitative change in the structure of dynamical fluctuations marked by the appearance of Dynamical Heterogeneities at the crossover temperature. On the other hand one can immediately see that the size of the peaks does not change significantly above and below $T_c$. In the following sections we will analyze the $B$-profile more carefully and we will see in particular that the correlation length is actually slightly decreasing below $T_c$.

\subsection{Viscosity and Diffusivity}

The viscosity $\eta$ and the diffusivity $D$ can be associated to different averages of the $B$-profile.
Following the same matching arguments valid in MCT we can associate the $\alpha$ time scale to the coefficient of $t^b$ in the total correlator $G(t)$, this leads naturally to:
\beq
\eta=\tau_\alpha \sim \left( {1 \over V}\int B(x) \, dx\right)^{-1/b}
\label{etaB}
\eeq
On the other hand, as discussed in \cite{Rizzo14b}, within SBR it is natural to consider a local relaxation time $\tau_\alpha(x) \sim B(x)^{-1/b}$ defined as the time scale where $-B(x)t^b$ is $O(1)$. One would therefore associate the diffusivity to the inverse of this local relaxation time:
\beq
D \sim  {1 \over V}\int B^{1/b}(x)\, dx
\label{DB}
\eeq

\begin{figure}
\begin{center}
\includegraphics[width=9cm]{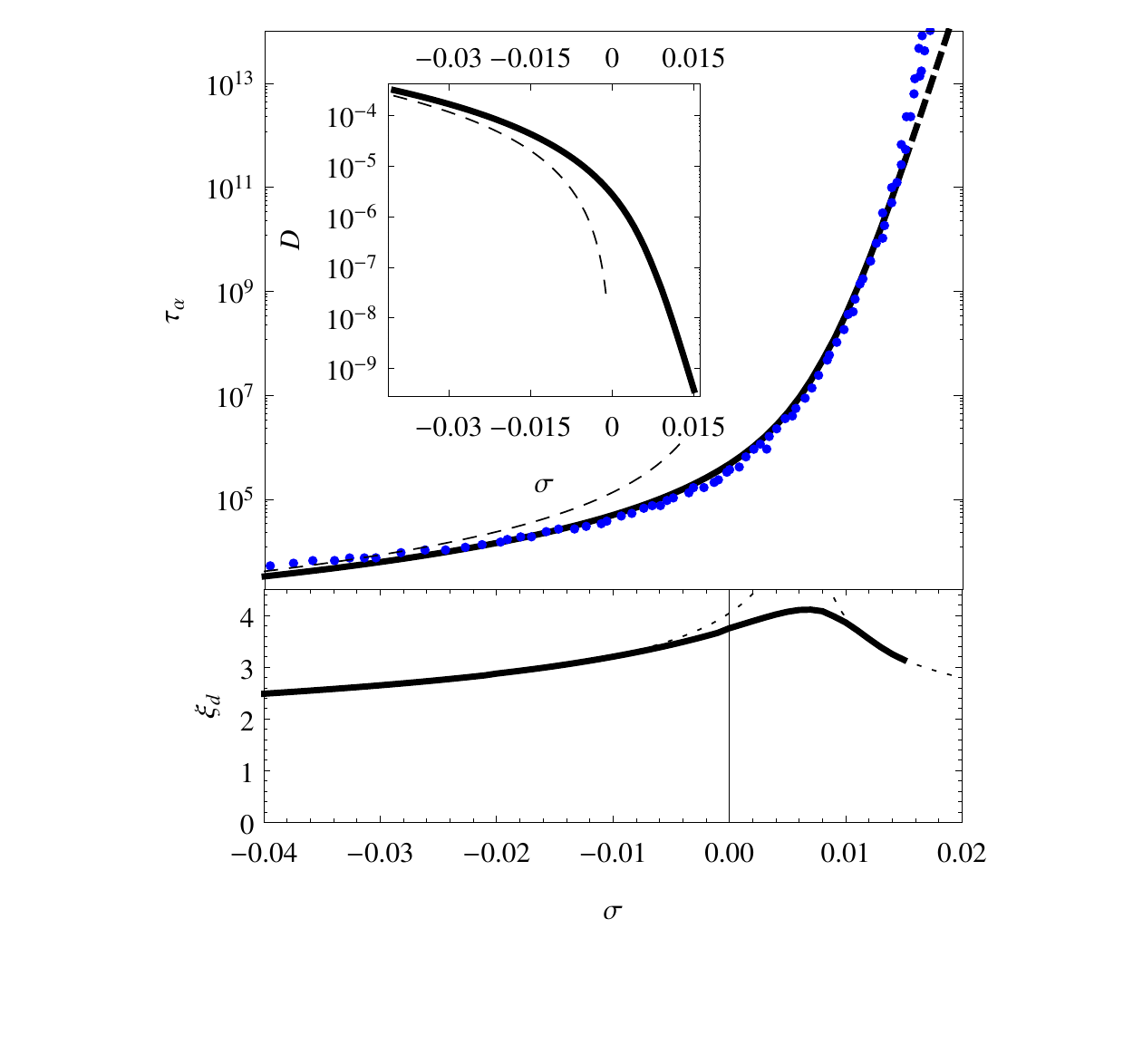}
\caption{Top: $\tau_\alpha$ vs. $\sigma$ from SBR in 3D. Inset: $D$ vs. $\sigma$ in 3D. Symbols are dielectric-spectroscopy
data for propylene carbonate from {\it Lunkenheimer et al} (2000). Dashed lines are MCT asymptotes. Bottom: The correlation length $\xi_d$ (defined as half-width at half-maximum of $\Gamma(r)$) as a function of $\sigma$.}
\label{fig:tDx}
\end{center}\end{figure}
In figure (\ref{fig:tDx}) we plot $\tau_\alpha$ and $D$ as a function of $\sigma$ together with exemplary experimental data, obtained by Schneider \textit{et~al.}\ \cite{Schneider99,Lunkenheimer00} from dielectric spectroscopy on propylene carbonate. 
The behavior of this quantities is similar to what has been found in the simplified model. Let us make a few comments on the figure:
\begin{itemize}

\item the divergence of $\tau_\alpha$ at $\sigma=0$ of ideal MCT is avoided but the growth rate changes from power-law to a more pronounced exponential-like growth: $T_c$ is avoided but marks a crossover.

\item The above property should be assessed remembering that there is no {\it ad hoc} assumption on activation processes in the derivation of SBR, rather the initial assumption is {\it i.e.} that $T_c$ marks a genuine phase transition.

\item The comparison with the data show that the description provided by SBR could work in an extended range of temperature, although it depends on a few number of parameters.

\item Used as fit functions the SBR expressions for $D$ and $\eta$ could help reconcile different estimates of $T_c$ based on asymptotic behavior.

\begin{figure}
\begin{center}
\includegraphics[width=6cm]{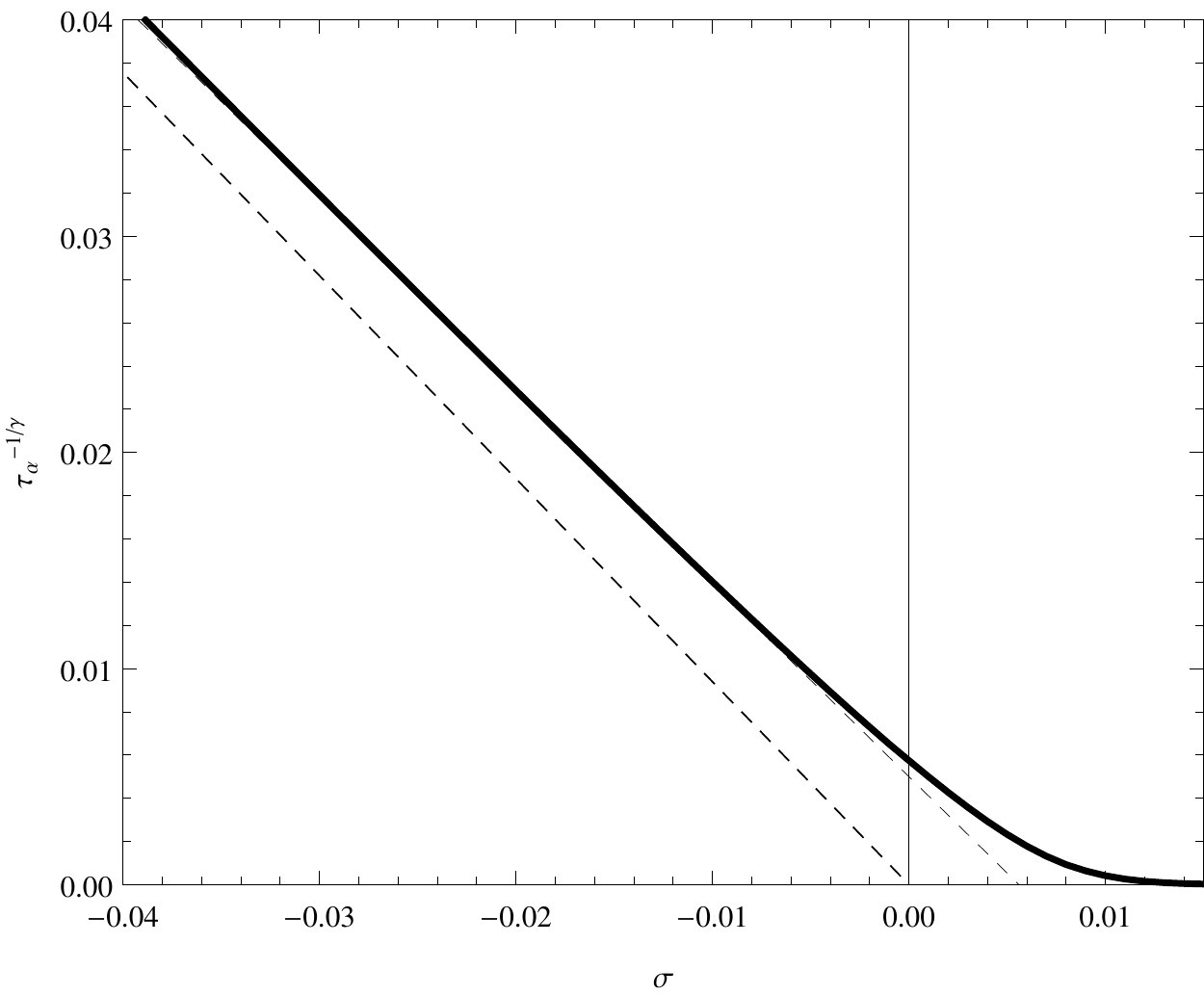}
\caption{Top: $\tau_\alpha^{-1/\gamma}$ vs. $\sigma$ from SBR in 3D. A power-law fit of the SBR expression would suggest a critical temperature definitively smaller than the actual value $\sigma=0$. Quantitatively the value of the shift depends on the parameters $\Delta\sigma$ and $\alpha$.}
\label{fig:lin}
\end{center}\end{figure}

\item SBR leads to an apparent shift of $T_c$ to lower values when fitted
with ideal MCT power laws and this could provide an explanation of the well-known shift of the quantitative prediction of $MCT$ with respect to simulation data \cite{Goetze09}, see fig. (\ref{fig:lin}).

\item In the figure the SBR expression for the $\tau_\alpha$ was used essentially as a fit function to the data but the actual computation of the parameters $\Delta\sigma$ and $\alpha$ is feasible for many systems that can be simulated numerically. These computations are left for future work.

\end{itemize}

As a technical remark we note that at given value of $\lambda$ the SBR equations depend in principle on three parameters $\alpha$, $\sigma$ and $\Delta \sigma$. However it suffices to solve it for fixed values of two of them. The natural way to do that is to consider fixed values of $\Delta \sigma$ and $\alpha$ and vary $\sigma$.
The general solution can then be obtained from the reference solution by means of appropriate rescalings. The numerical data shown here correspond to the following choice of the parameters:
\beq
\alpha=.2\,,\, \Delta \sigma=.1,\, \lambda=.75\ ,
\eeq

\subsection{Non-monotonous Correlation Length}

The spatial fluctuations of the dynamics in the late $\beta$ regime are conveniently encoded by the spatial fluctuations of the $B$-profile. 
We introduce the self-correlation of the $B$-profile as:
\beq
\Gamma(r) \equiv  {1 \over V}\int B(x) B(x+r)\, dx
\eeq
The numerical solution shows that while the absolute value of $\Gamma(r)$ changes by orders of magnitude upon crossing $\sigma=0$ (it is related to $\tau_\alpha$), its shape and length-scale vary much less.
In order to focus solely on the space dependence we introduce the normalized self-correlation:
\beq
\Gamma_n(r) \equiv \frac{\Gamma(r)-\Gamma(\infty)}{\Gamma(0)-\Gamma(\infty)}
\eeq
We observe that in 3D  $\Gamma_n(r)$ has a bell-shaped form with rapidly decaying tails both below and above $\sigma=0$. Quite interestingly it turns out that the width of $\Gamma_n(r)$ has a non-monotonous behavior with $\sigma$: in the bottom of figure (\ref{fig:tDx}) we plot the half-width at half-maximum of $\Gamma_n(r)$ as a function of $\sigma$ and use it as a definition of the dynamical correlation length. The figure demonstrates what we anticipated in the introduction:
the dynamical correlation length increases approaching $\sigma=0$ from the liquid phase, it saturates to a maximum at $\sigma=\sigma_{max}$  slightly below $T_c$  and then it decreases again. In the same region the relaxation time increases instead by orders of magnitude implying that near $\sigma=0$ {\it the correlation length decouples from the relaxation time}.
For instance considering the interval $\sigma=[-.01,.015]$ the correlation length increases and decreases again with an approximate $25\%$ excursion while the relaxation time increases by almost $8$ orders of magnitude.

It is tempting to put these results in connection with recent observations of non-monotonous correlation lengths in numerical simulations \cite{Kob12} and experiments \cite{Nagamanasa14}. These observations however should be contrasted with recent measurements  \cite{Flenner12,Flenner13} of dynamical correlation lengths for the same systems obtained with the more standard methods of Refs. \cite{Franz99,Donati02,Lacevic02,Lacevic03}. In this case a monotonous behavior was instead observed, albeit displaying evidences of saturation towards a maximum \cite{Flenner12,Flenner13}. With regard to this open issue we note that, as discussed in \cite{Rizzo14}, SBR is valid as it is near the crossover region. In particular the assumption that the separation parameter $\sigma$ is the sole temperature-dependent quantity (a typical assumption for a genuine phase transition) could be too strong well below $T_c$. 
On the other hand according to fig. (\ref{fig:tDx}) the correlation length does not change too much with the temperature, and one cannot exclude the possibility that the temperature dependence of the parameters $\alpha$ and $\Delta\sigma$  alters the non-monotonous behavior in actual systems. Qualitatively however the scenario of fig. (\ref{fig:B-slices}) would remain the same: the amplitude of fluctuations (the height of the peaks)
increase considerably while their size does not change significantly and this should be considered the essential feature of SBR independently on weather this size is actually increasing or decreasing.

\subsection{Dynamical Heterogeneitites}

Given that the correlation length displays a symmetrically decrease far away from $T_c$ both  above and below one may ask what is actually happening  at the crossover. 
As we saw before the answer is that there is a dramatic change in the structure of dynamical fluctuations above and below the crossover temperature as can be seen by direct inspection of the $B$-profiles. Let us now examine figure (\ref{fig:B-slices}) thoroughly.
We can distinguish two regimes above and below the crossover temperature. 
Above the critical temperature ($\sigma<0$) the normalized $B$-profile has fluctuations of order $O(1)$ around its average value which is one by definition.
The profile is characterized by  peaks whose width is of the order of magnitude of the correlation length. Increasing $\sigma$ towards $\sigma=0$ the profile does not change much, although we observe slightly larger fluctuations in the height of the peaks. In this regime the profile evolves much as in a second-order phase transition: the correlation length increases and so do the amplitude fluctuations. Here the correlation length carries relevant information: the system is essentially scale-invariant in the sense that the profiles at different values of $\sigma \leq 0$ look the same once space is rescaled proportionally to the correlation length.
However at higher values of $\sigma$ ($\sigma=.006$) the $B$-profile starts to change qualitatively: the height of the {\it typical} region decreases as more and more of the lowest peaks disappear, conversely the few peaks left increase their relative height because they carry all the weight. 
As a result at $\sigma=.012$  we have a completely different landscape characterized by rare regions where the dynamics is relatively much faster ($B(x)/\overline{B} \gg 1$ at the peaks) then the surrounding slowly-moving regions ($B(x)/\overline{B} \ll 1$ in the typical region). 
\comment{\begin{figure}
\begin{center}
\includegraphics[width=8.5cm]{B-line.pdf}
\caption{Log-plot of the Normalized $B$-profile $B(x)/\overline{B}$ on the line $x=20$ cut from the planes of figures \ref{fig:B-slices} (increasing values of $\sigma$ correspond to increasing line thickness).}
\label{fig:B-line}
\end{center}\end{figure}}
Quantitatively we see that while fluctuations are $O(1)$ at $\sigma=-.006$, for $\sigma=0.012$ the normalized $B$-profile is $2-3$ orders of magnitude larger in the rare fast regions with respect to the neighboring slow regions.
From figure (\ref{fig:B-slices}) we also see that the size and shape of the peaks does not change much, which result in the fact that the correlation length does not change significantly for the values of $\sigma$ considered (as seen in figure (\ref{fig:tDx})).

These features suggest that we should consider carefully the role of the correlation length at the glass crossover and the very same notion of Dynamical Heterogeneity (DH).
Although dynamical fluctuations grow in both regimes they have power-law increase and remains relatively small in the first regime while they grow significantly in the second regime where they have essentially an activated nature. Therefore only in the second regime we should actually talk of Dynamical Heterogeneities as defined from experiments \cite{Ediger00} and as a consequence we should conclude that within SBR {\it DH are not intrinsically associated with an increasing correlation length}.

\begin{figure}
\begin{center}
\includegraphics[width=8.5cm]{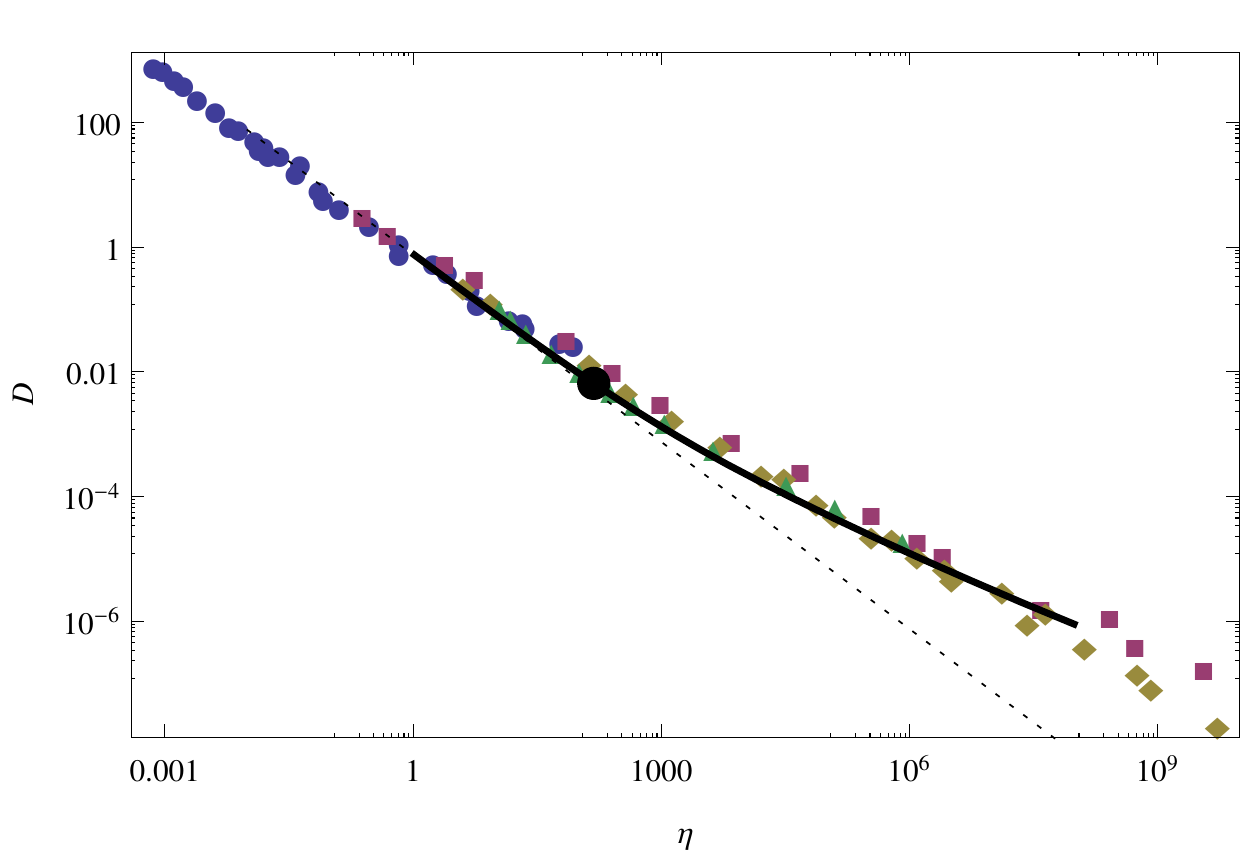}
\caption{Viscosity-Diffusivity parametric plot from SBR in 3D.
Dashed line indicates $D\propto\eta^{-1}$ (Stokes-Einstein
relation), a large circle marks $T=T_c$.
Smaller symbols: experimental data for o-terphenyl from Ref.~\cite{Lohfink92} (circles: tracer diffusion at $T\gtrsim T_c$; squares: diffusion of flourescent ACR dye; diamonds: TTI dye); simulation results for a harmonic-sphere mixture (triangles, Ref.~\cite{Flenner13}).
}
\label{fig:ser3D}
\end{center}\end{figure}

It is to be expected that the qualitative change in the structure of the dynamical fluctuations from scale-invariant-like to activated-like can be detected by comparing observables that are associated to different averages of $B(x)$. In figure (\ref{fig:ser3D}) we plot parametrically the viscosity and the diffusivity computed according to expressions (\ref{etaB}) and (\ref{DB}). They obey the Stokes-Einstein relationship (SER) $D \eta \propto 1$ for temperatures above $T_c$, while below the crossover (the black dot) there is  a violation of the SER. Once again this is a typical feature of glassy systems as the exemplary data demonstrates.  

\section{Discussion}

In order to assess the properties of SBR described before one should bare in mind that it is not a phenomenological theory and there is instead a non-trivial mathematical connection between MCT and similar microscopic theories characterized by an ideal glass transition. As we said in the introduction this connection is rather unexpected because it is based on the application of perturbative methods to an effective dynamical field theory which coincides at the tree level with the equation for the critical correlator of ideal MCT.

The dynamical field theory from which SBR was derived is closely related to a {\it static} replicated field theory that was associated to the glass problem long ago \cite{Kirkpatrick87b,Kirkpatrick87c,Kirkpatrick87}. In particular a {\it static} stochastic equation was first derived from the replicated field theory in \cite{Franz11}. One should be aware that this is the same (static) field theory that lies at the heart of the Random-Fisrt-Order-Transition (RFOT) theory \cite{Kirkpatrick89}. As it is well known, RFOT  claims to include the early stage of vitrification described by MCT phenomenology but is definitively more focused on the existence of an ideal glass transition below the calorimetric glass transition $T_g$ and put emphasis on the fact that the corresponding dynamical slowing down is accompanied by a diverging correlation length. 
SBR is instead a theory of the glass crossover and it is not clear {\it a priori} how deep in the super-cooled regime it provides a good description. Thus the two theories are not necessarily incompatible because they describe different temperature regimes. On the other hand clearly if one were to speculate on the deeply supercooled regime starting solely from the SBR description of the crossover one would think of a scenario in which the correlation length does not play a crucial role and we will further comment on this issue in the final paragraphs.

\begin{figure}
\includegraphics[width=.9\linewidth]{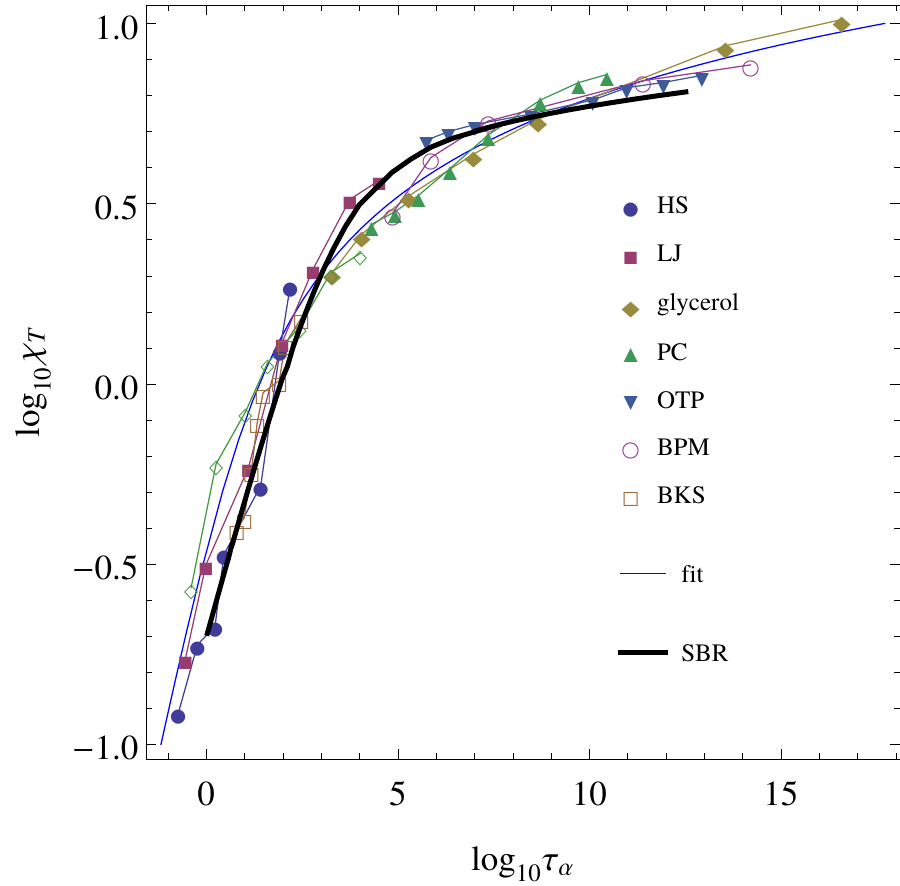}
\caption{Peak of the thermal susceptibility $\chi_T$ as a function
of relaxation time $\tau_\alpha$ from SBR theory in 3D (solid line);  Symbols: experimental data plotted as in Ref.~\protect\cite{DalleFerrier07} (circles: Lennard Jones mixture; squares: hard spheres; BKS silica: diamonds; triangles: propylene carbonate; inverted triangles: glycerol; open circles: OTP; open squares: salol). Thin line: expression proposed in Ref.~\protect\cite{DalleFerrier07}.}
\label{fig:XT-vs-ta}
\end{figure}

In this respect it is interesting to consider the behavior of the  susceptibility with respect to external parameters. Currently there are great efforts to measure this quantity directly  \cite{Berthier05,Thibierge10,Bauer13} the main motivation being the proposed existence of
a direct connection between $\chi_T$ and the size of dynamical heterogeneities $N_{\mathrm corr}$ and thus inferring a monotonous increase of the latter.
As discussed in \cite{Rizzo14b} within SBR it is natural to associate these susceptibilities to $d \ln \tau_\alpha / d\sigma$, the resulting plot obtained from the 3D data is shown in fig. (\ref{fig:XT-vs-ta}) and is similar qualitatively to the result of the simplified model. However within SBR the increase of dynamical susceptibilities  is accompanied in the super-cooled region by a decrease of the correlation length and this questions  the interpretation of the experimental data as evidence of an increase of $N_{\mathrm corr}$.

From fig. (\ref{fig:B-slices}) we see that in the supercooled regime fluctuations of the $B$-profile increase constantly in amplitude (not in size).
We should notice that the theory however cannot sustain too large fluctuations and must be abandoned at some point. The reason is that the time when the total correlator $G(t)$ (controlled by $\overline{B}$) enters the $\alpha$-regime becomes much larger than the time when  the fast regions (corresponding to $B(x)/\overline{B} \gg 1$) have entered the $\alpha$ regime. On the other hand as soon as the fast regions enter the $\alpha$ regime the description should be abandoned because $g(x,t)$ is large (although locally) and the scaling condition $g(x,t) \ll 1$ for the validity of the critical MCT equation and correspondingly SBR ceases to be valid. In particular on time scale where $G(t)$ becomes $O(1)$ and negative the fast regions would have a local $g(x,t)$  negative and very large in absolute value, but this cannot happens because the density-density correlator ${\bf F}^c(q)+g(x,t) {\bf H}(q)$ must remain positive everywhere. What happens when fluctuations become too large cannot be predicted from SBR. One possibility is that the description provided by SBR remains valid because fluctuations are somehow damped by quantitative corrections. Indeed not only the separation parameter but also the coupling constants drift with the external parameters and this changes the amplitude and extension of the fluctuations. Another possibility is that one enters a full-fledged activated regime and it is tempting to make some conjectures on it inspired by SBR.
In particular the fact that in SBR the correlation length remains relatively large, albeit decreasing, below $T_c$ suggest that this activated regime could display {\it important differences with respect to ordinary activated dynamics},
 which is driven by microscopic ({\it ie.} at the single particle scale) events occurring exponentially rarely in time but lasting for microscopic times.
In this non-standard activated pictures the elementary events (of which the peaks would be the precursors) are non-standard in the sense that they involve a relatively large number of particles (although not diverging but actually decreasing) and have an intrinsic time-scale considerably larger than the microscopic one.

\bibliographystyle{apsrev4-1}
\bibliography{lit}

\begin{thebibliography}{27}%
\makeatletter
\providecommand \@ifxundefined [1]{%
 \@ifx{#1\undefined}
}%
\providecommand \@ifnum [1]{%
 \ifnum #1\expandafter \@firstoftwo
 \else \expandafter \@secondoftwo
 \fi
}%
\providecommand \@ifx [1]{%
 \ifx #1\expandafter \@firstoftwo
 \else \expandafter \@secondoftwo
 \fi
}%
\providecommand \natexlab [1]{#1}%
\providecommand \enquote  [1]{``#1''}%
\providecommand \bibnamefont  [1]{#1}%
\providecommand \bibfnamefont [1]{#1}%
\providecommand \citenamefont [1]{#1}%
\providecommand \href@noop [0]{\@secondoftwo}%
\providecommand \href [0]{\begingroup \@sanitize@url \@href}%
\providecommand \@href[1]{\@@startlink{#1}\@@href}%
\providecommand \@@href[1]{\endgroup#1\@@endlink}%
\providecommand \@sanitize@url [0]{\catcode `\\12\catcode `\$12\catcode
  `\&12\catcode `\#12\catcode `\^12\catcode `\_12\catcode `\%12\relax}%
\providecommand \@@startlink[1]{}%
\providecommand \@@endlink[0]{}%
\providecommand \url  [0]{\begingroup\@sanitize@url \@url }%
\providecommand \@url [1]{\endgroup\@href {#1}{\urlprefix }}%
\providecommand \urlprefix  [0]{URL }%
\providecommand \Eprint [0]{\href }%
\providecommand \doibase [0]{http://dx.doi.org/}%
\providecommand \selectlanguage [0]{\@gobble}%
\providecommand \bibinfo  [0]{\@secondoftwo}%
\providecommand \bibfield  [0]{\@secondoftwo}%
\providecommand \translation [1]{[#1]}%
\providecommand \BibitemOpen [0]{}%
\providecommand \bibitemStop [0]{}%
\providecommand \bibitemNoStop [0]{.\EOS\space}%
\providecommand \EOS [0]{\spacefactor3000\relax}%
\providecommand \BibitemShut  [1]{\csname bibitem#1\endcsname}%
\let\auto@bib@innerbib\@empty
%</preamble>
\bibitem [{\citenamefont {G\"otze}(2009)}]{Goetze09}%
  \BibitemOpen
  \bibfield  {author} {\bibinfo {author} {\bibfnamefont {W.}~\bibnamefont
  {G\"otze}},\ }\href@noop {} {\emph {\bibinfo {title} {Complex Dynamics of
  Glass-Forming Liquids}}}\ (\bibinfo  {publisher} {Oxford University Press},\
  \bibinfo {address} {Oxford},\ \bibinfo {year} {2009})\BibitemShut {NoStop}%
\bibitem [{\citenamefont {Rizzo}(2014)}]{Rizzo14}%
  \BibitemOpen
  \bibfield  {author} {\bibinfo {author} {\bibfnamefont {T.}~\bibnamefont
  {Rizzo}},\ }\href {http://stacks.iop.org/0295-5075/106/i=5/a=56003}
  {\bibfield  {journal} {\bibinfo  {journal} {EPL (Europhysics Letters)}\
  }\textbf {\bibinfo {volume} {106}},\ \bibinfo {pages} {56003} (\bibinfo
  {year} {2014})}\BibitemShut {NoStop}%
\bibitem [{\citenamefont {Rizzo}\ and\ \citenamefont
  {Voigtmann}(2014)}]{Rizzo14b}%
  \BibitemOpen
  \bibfield  {author} {\bibinfo {author} {\bibfnamefont {T.}~\bibnamefont
  {Rizzo}}\ and\ \bibinfo {author} {\bibfnamefont {T.}~\bibnamefont
  {Voigtmann}},\ }\href@noop {} {} (\bibinfo {year} {2014}),\ \Eprint
  {http://arxiv.org/abs/1403.2764} {arXiv:1403.2764} \BibitemShut {NoStop}%
\bibitem [{\citenamefont {Biroli}\ \emph {et~al.}(2006)\citenamefont {Biroli},
  \citenamefont {Bouchaud}, \citenamefont {Miyazaki},\ and\ \citenamefont
  {Reichman}}]{Biroli06}%
  \BibitemOpen
  \bibfield  {author} {\bibinfo {author} {\bibfnamefont {G.}~\bibnamefont
  {Biroli}}, \bibinfo {author} {\bibfnamefont {J.-P.}\ \bibnamefont
  {Bouchaud}}, \bibinfo {author} {\bibfnamefont {K.}~\bibnamefont {Miyazaki}},
  \ and\ \bibinfo {author} {\bibfnamefont {D.~R.}\ \bibnamefont {Reichman}},\
  }\href@noop {} {\bibfield  {journal} {\bibinfo  {journal} {Phys. Rev. Lett.}\
  }\textbf {\bibinfo {volume} {97}},\ \bibinfo {pages} {195701} (\bibinfo
  {year} {2006})}\BibitemShut {NoStop}%
\bibitem [{\citenamefont {Franz}\ \emph {et~al.}(2012)\citenamefont {Franz},
  \citenamefont {Jacquin}, \citenamefont {Parisi}, \citenamefont {Urbani},\
  and\ \citenamefont {Zamponi}}]{Franz12}%
  \BibitemOpen
  \bibfield  {author} {\bibinfo {author} {\bibfnamefont {S.}~\bibnamefont
  {Franz}}, \bibinfo {author} {\bibfnamefont {H.}~\bibnamefont {Jacquin}},
  \bibinfo {author} {\bibfnamefont {G.}~\bibnamefont {Parisi}}, \bibinfo
  {author} {\bibfnamefont {P.}~\bibnamefont {Urbani}}, \ and\ \bibinfo {author}
  {\bibfnamefont {F.}~\bibnamefont {Zamponi}},\ }\href {\doibase
  10.1073/pnas.1216578109} {\bibfield  {journal} {\bibinfo  {journal}
  {Proceedings of the National Academy of Sciences}\ }\textbf {\bibinfo
  {volume} {109}},\ \bibinfo {pages} {18725} (\bibinfo {year} {2012})},\
  \Eprint
  {http://arxiv.org/abs/http://www.pnas.org/content/109/46/18725.full.pdf+html}
  {http://www.pnas.org/content/109/46/18725.full.pdf+html} \BibitemShut
  {NoStop}%
\bibitem [{\citenamefont {Rizzo}(2013)}]{Rizzo13}%
  \BibitemOpen
  \bibfield  {author} {\bibinfo {author} {\bibfnamefont {T.}~\bibnamefont
  {Rizzo}},\ }\href@noop {} {} (\bibinfo {year} {2013}),\ \Eprint
  {http://arxiv.org/abs/1307.4303} {arXiv:1307.4303} \BibitemShut {NoStop}%
\bibitem [{\citenamefont {Schneider}\ \emph {et~al.}(1999)\citenamefont
  {Schneider}, \citenamefont {Lunkenheimer}, \citenamefont {Brand},\ and\
  \citenamefont {Loidl}}]{Schneider99}%
  \BibitemOpen
  \bibfield  {author} {\bibinfo {author} {\bibfnamefont {U.}~\bibnamefont
  {Schneider}}, \bibinfo {author} {\bibfnamefont {P.}~\bibnamefont
  {Lunkenheimer}}, \bibinfo {author} {\bibfnamefont {R.}~\bibnamefont {Brand}},
  \ and\ \bibinfo {author} {\bibfnamefont {A.}~\bibnamefont {Loidl}},\
  }\href@noop {} {\bibfield  {journal} {\bibinfo  {journal} {Phys. Rev. E}\
  }\textbf {\bibinfo {volume} {59}},\ \bibinfo {pages} {6924} (\bibinfo {year}
  {1999})}\BibitemShut {NoStop}%
\bibitem [{\citenamefont {Lunkenheimer}\ \emph {et~al.}(2000)\citenamefont
  {Lunkenheimer}, \citenamefont {Schneider}, \citenamefont {Brand},\ and\
  \citenamefont {Loidl}}]{Lunkenheimer00}%
  \BibitemOpen
  \bibfield  {author} {\bibinfo {author} {\bibfnamefont {P.}~\bibnamefont
  {Lunkenheimer}}, \bibinfo {author} {\bibfnamefont {U.}~\bibnamefont
  {Schneider}}, \bibinfo {author} {\bibfnamefont {R.}~\bibnamefont {Brand}}, \
  and\ \bibinfo {author} {\bibfnamefont {A.}~\bibnamefont {Loidl}},\
  }\href@noop {} {\bibfield  {journal} {\bibinfo  {journal} {Contemp. Phys.}\
  }\textbf {\bibinfo {volume} {41}},\ \bibinfo {pages} {15} (\bibinfo {year}
  {2000})}\BibitemShut {NoStop}%
\bibitem [{\citenamefont {Kob}\ \emph {et~al.}(2012)\citenamefont {Kob},
  \citenamefont {Rold{\'a}n-Vargas},\ and\ \citenamefont {Berthier}}]{Kob12}%
  \BibitemOpen
  \bibfield  {author} {\bibinfo {author} {\bibfnamefont {W.}~\bibnamefont
  {Kob}}, \bibinfo {author} {\bibfnamefont {S.}~\bibnamefont
  {Rold{\'a}n-Vargas}}, \ and\ \bibinfo {author} {\bibfnamefont
  {L.}~\bibnamefont {Berthier}},\ }\href@noop {} {\bibfield  {journal}
  {\bibinfo  {journal} {Nature Physics}\ }\textbf {\bibinfo {volume} {8}},\
  \bibinfo {pages} {164} (\bibinfo {year} {2012})}\BibitemShut {NoStop}%
\bibitem [{\citenamefont {Nagamanasa}\ \emph {et~al.}(2014)\citenamefont
  {Nagamanasa}, \citenamefont {Gokhale}, \citenamefont {Sood},\ and\
  \citenamefont {Ganapathy}}]{Nagamanasa14}%
  \BibitemOpen
  \bibfield  {author} {\bibinfo {author} {\bibfnamefont {K.~H.}\ \bibnamefont
  {Nagamanasa}}, \bibinfo {author} {\bibfnamefont {S.}~\bibnamefont {Gokhale}},
  \bibinfo {author} {\bibfnamefont {A.}~\bibnamefont {Sood}}, \ and\ \bibinfo
  {author} {\bibfnamefont {R.}~\bibnamefont {Ganapathy}},\ }\href@noop {}
  {\bibfield  {journal} {\bibinfo  {journal} {arXiv preprint arXiv:1408.5485}\
  } (\bibinfo {year} {2014})}\BibitemShut {NoStop}%
\bibitem [{\citenamefont {Flenner}\ and\ \citenamefont
  {Szamel}(2012)}]{Flenner12}%
  \BibitemOpen
  \bibfield  {author} {\bibinfo {author} {\bibfnamefont {E.}~\bibnamefont
  {Flenner}}\ and\ \bibinfo {author} {\bibfnamefont {G.}~\bibnamefont
  {Szamel}},\ }\href@noop {} {\bibfield  {journal} {\bibinfo  {journal} {Nature
  Physics}\ }\textbf {\bibinfo {volume} {8}},\ \bibinfo {pages} {696} (\bibinfo
  {year} {2012})}\BibitemShut {NoStop}%
\bibitem [{\citenamefont {Flenner}\ and\ \citenamefont
  {Szamel}(2013)}]{Flenner13}%
  \BibitemOpen
  \bibfield  {author} {\bibinfo {author} {\bibfnamefont {E.}~\bibnamefont
  {Flenner}}\ and\ \bibinfo {author} {\bibfnamefont {G.}~\bibnamefont
  {Szamel}},\ }\href@noop {} {\bibfield  {journal} {\bibinfo  {journal} {J.
  Chem. Phys.}\ }\textbf {\bibinfo {volume} {138}},\ \bibinfo {pages} {12A523}
  (\bibinfo {year} {2013})}\BibitemShut {NoStop}%
\bibitem [{\citenamefont {Franz}\ \emph {et~al.}(1999)\citenamefont {Franz},
  \citenamefont {Donati}, \citenamefont {Parisi},\ and\ \citenamefont
  {Glotzer}}]{Franz99}%
  \BibitemOpen
  \bibfield  {author} {\bibinfo {author} {\bibfnamefont {S.}~\bibnamefont
  {Franz}}, \bibinfo {author} {\bibfnamefont {C.}~\bibnamefont {Donati}},
  \bibinfo {author} {\bibfnamefont {G.}~\bibnamefont {Parisi}}, \ and\ \bibinfo
  {author} {\bibfnamefont {S.~C.}\ \bibnamefont {Glotzer}},\ }\href@noop {}
  {\bibfield  {journal} {\bibinfo  {journal} {Philosophical Magazine B}\
  }\textbf {\bibinfo {volume} {79}},\ \bibinfo {pages} {1827} (\bibinfo {year}
  {1999})}\BibitemShut {NoStop}%
\bibitem [{\citenamefont {Donati}\ \emph {et~al.}(2002)\citenamefont {Donati},
  \citenamefont {Franz}, \citenamefont {Glotzer},\ and\ \citenamefont
  {Parisi}}]{Donati02}%
  \BibitemOpen
  \bibfield  {author} {\bibinfo {author} {\bibfnamefont {C.}~\bibnamefont
  {Donati}}, \bibinfo {author} {\bibfnamefont {S.}~\bibnamefont {Franz}},
  \bibinfo {author} {\bibfnamefont {S.~C.}\ \bibnamefont {Glotzer}}, \ and\
  \bibinfo {author} {\bibfnamefont {G.}~\bibnamefont {Parisi}},\ }\href@noop {}
  {\bibfield  {journal} {\bibinfo  {journal} {Journal of non-crystalline
  solids}\ }\textbf {\bibinfo {volume} {307}},\ \bibinfo {pages} {215}
  (\bibinfo {year} {2002})}\BibitemShut {NoStop}%
\bibitem [{\citenamefont {La{\v{c}}evi{\'c}}\ \emph {et~al.}(2002)\citenamefont
  {La{\v{c}}evi{\'c}}, \citenamefont {Starr}, \citenamefont {Schr{\o}der},
  \citenamefont {Novikov},\ and\ \citenamefont {Glotzer}}]{Lacevic02}%
  \BibitemOpen
  \bibfield  {author} {\bibinfo {author} {\bibfnamefont {N.}~\bibnamefont
  {La{\v{c}}evi{\'c}}}, \bibinfo {author} {\bibfnamefont {F.~W.}\ \bibnamefont
  {Starr}}, \bibinfo {author} {\bibfnamefont {T.}~\bibnamefont {Schr{\o}der}},
  \bibinfo {author} {\bibfnamefont {V.}~\bibnamefont {Novikov}}, \ and\
  \bibinfo {author} {\bibfnamefont {S.}~\bibnamefont {Glotzer}},\ }\href@noop
  {} {\bibfield  {journal} {\bibinfo  {journal} {Physical Review E}\ }\textbf
  {\bibinfo {volume} {66}},\ \bibinfo {pages} {030101} (\bibinfo {year}
  {2002})}\BibitemShut {NoStop}%
\bibitem [{\citenamefont {La{\v{c}}evi{\'c}}\ \emph {et~al.}(2003)\citenamefont
  {La{\v{c}}evi{\'c}}, \citenamefont {Starr}, \citenamefont {Schr{\o}der},\
  and\ \citenamefont {Glotzer}}]{Lacevic03}%
  \BibitemOpen
  \bibfield  {author} {\bibinfo {author} {\bibfnamefont {N.}~\bibnamefont
  {La{\v{c}}evi{\'c}}}, \bibinfo {author} {\bibfnamefont {F.~W.}\ \bibnamefont
  {Starr}}, \bibinfo {author} {\bibfnamefont {T.}~\bibnamefont {Schr{\o}der}},
  \ and\ \bibinfo {author} {\bibfnamefont {S.}~\bibnamefont {Glotzer}},\
  }\href@noop {} {\bibfield  {journal} {\bibinfo  {journal} {The Journal of
  chemical physics}\ }\textbf {\bibinfo {volume} {119}},\ \bibinfo {pages}
  {7372} (\bibinfo {year} {2003})}\BibitemShut {NoStop}%
\bibitem [{\citenamefont {Ediger}(2000)}]{Ediger00}%
  \BibitemOpen
  \bibfield  {author} {\bibinfo {author} {\bibfnamefont {M.~D.}\ \bibnamefont
  {Ediger}},\ }\href@noop {} {\bibfield  {journal} {\bibinfo  {journal} {Annu.
  Rev. Phys. Chem.}\ }\textbf {\bibinfo {volume} {51}},\ \bibinfo {pages} {99}
  (\bibinfo {year} {2000})}\BibitemShut {NoStop}%
\bibitem [{\citenamefont {Lohfink}\ and\ \citenamefont
  {Sillescu}(1992)}]{Lohfink92}%
  \BibitemOpen
  \bibfield  {author} {\bibinfo {author} {\bibfnamefont {M.}~\bibnamefont
  {Lohfink}}\ and\ \bibinfo {author} {\bibfnamefont {H.}~\bibnamefont
  {Sillescu}},\ }\href@noop {} {\bibfield  {journal} {\bibinfo  {journal} {AIP
  Conf. Proc.}\ }\textbf {\bibinfo {volume} {256}},\ \bibinfo {pages} {30}
  (\bibinfo {year} {1992})}\BibitemShut {NoStop}%
\bibitem [{\citenamefont {Kirkpatrick}\ and\ \citenamefont
  {Thirumalai}(1987{\natexlab{a}})}]{Kirkpatrick87b}%
  \BibitemOpen
  \bibfield  {author} {\bibinfo {author} {\bibfnamefont {T.~R.}\ \bibnamefont
  {Kirkpatrick}}\ and\ \bibinfo {author} {\bibfnamefont {D.}~\bibnamefont
  {Thirumalai}},\ }\href {\doibase 10.1103/PhysRevB.36.5388} {\bibfield
  {journal} {\bibinfo  {journal} {Phys. Rev. B}\ }\textbf {\bibinfo {volume}
  {36}},\ \bibinfo {pages} {5388} (\bibinfo {year}
  {1987}{\natexlab{a}})}\BibitemShut {NoStop}%
\bibitem [{\citenamefont {Kirkpatrick}\ and\ \citenamefont
  {Thirumalai}(1987{\natexlab{b}})}]{Kirkpatrick87c}%
  \BibitemOpen
  \bibfield  {author} {\bibinfo {author} {\bibfnamefont {T.~R.}\ \bibnamefont
  {Kirkpatrick}}\ and\ \bibinfo {author} {\bibfnamefont {D.}~\bibnamefont
  {Thirumalai}},\ }\href {\doibase 10.1103/PhysRevLett.58.2091} {\bibfield
  {journal} {\bibinfo  {journal} {Phys. Rev. Lett.}\ }\textbf {\bibinfo
  {volume} {58}},\ \bibinfo {pages} {2091} (\bibinfo {year}
  {1987}{\natexlab{b}})}\BibitemShut {NoStop}%
\bibitem [{\citenamefont {Kirkpatrick}\ and\ \citenamefont
  {Wolynes}(1987)}]{Kirkpatrick87}%
  \BibitemOpen
  \bibfield  {author} {\bibinfo {author} {\bibfnamefont {T.~R.}\ \bibnamefont
  {Kirkpatrick}}\ and\ \bibinfo {author} {\bibfnamefont {P.~G.}\ \bibnamefont
  {Wolynes}},\ }\href@noop {} {\bibfield  {journal} {\bibinfo  {journal} {Phys.
  Rev. B}\ }\textbf {\bibinfo {volume} {36}},\ \bibinfo {pages} {8552}
  (\bibinfo {year} {1987})}\BibitemShut {NoStop}%
\bibitem [{\citenamefont {Franz}\ \emph {et~al.}(2011)\citenamefont {Franz},
  \citenamefont {Parisi}, \citenamefont {Ricci-Tersenghi},\ and\ \citenamefont
  {Rizzo}}]{Franz11}%
  \BibitemOpen
  \bibfield  {author} {\bibinfo {author} {\bibfnamefont {S.}~\bibnamefont
  {Franz}}, \bibinfo {author} {\bibfnamefont {G.}~\bibnamefont {Parisi}},
  \bibinfo {author} {\bibfnamefont {F.}~\bibnamefont {Ricci-Tersenghi}}, \ and\
  \bibinfo {author} {\bibfnamefont {T.}~\bibnamefont {Rizzo}},\ }\href
  {\doibase 10.1140/epje/i2011-11102-0} {\bibfield  {journal} {\bibinfo
  {journal} {The European Physical Journal E}\ }\textbf {\bibinfo {volume}
  {34}},\ \bibinfo {pages} {1} (\bibinfo {year} {2011})}\BibitemShut {NoStop}%
\bibitem [{\citenamefont {Kirkpatrick}\ \emph {et~al.}(1989)\citenamefont
  {Kirkpatrick}, \citenamefont {Thirumalai},\ and\ \citenamefont
  {Wolynes}}]{Kirkpatrick89}%
  \BibitemOpen
  \bibfield  {author} {\bibinfo {author} {\bibfnamefont {T.~R.}\ \bibnamefont
  {Kirkpatrick}}, \bibinfo {author} {\bibfnamefont {D.}~\bibnamefont
  {Thirumalai}}, \ and\ \bibinfo {author} {\bibfnamefont {P.~G.}\ \bibnamefont
  {Wolynes}},\ }\href {\doibase 10.1103/PhysRevA.40.1045} {\bibfield  {journal}
  {\bibinfo  {journal} {Phys. Rev. A}\ }\textbf {\bibinfo {volume} {40}},\
  \bibinfo {pages} {1045} (\bibinfo {year} {1989})}\BibitemShut {NoStop}%
\bibitem [{\citenamefont {Dalle-Ferrier}\ \emph {et~al.}(2007)\citenamefont
  {Dalle-Ferrier}, \citenamefont {Thibierge}, \citenamefont {Alba-Simionesco},
  \citenamefont {Berthier}, \citenamefont {Biroli}, \citenamefont {Bouchaud},
  \citenamefont {Ladieu}, \citenamefont {L'H\^ote},\ and\ \citenamefont
  {Tarjus}}]{DalleFerrier07}%
  \BibitemOpen
  \bibfield  {author} {\bibinfo {author} {\bibfnamefont {C.}~\bibnamefont
  {Dalle-Ferrier}}, \bibinfo {author} {\bibfnamefont {C.}~\bibnamefont
  {Thibierge}}, \bibinfo {author} {\bibfnamefont {C.}~\bibnamefont
  {Alba-Simionesco}}, \bibinfo {author} {\bibfnamefont {L.}~\bibnamefont
  {Berthier}}, \bibinfo {author} {\bibfnamefont {G.}~\bibnamefont {Biroli}},
  \bibinfo {author} {\bibfnamefont {J.-P.}\ \bibnamefont {Bouchaud}}, \bibinfo
  {author} {\bibfnamefont {F.}~\bibnamefont {Ladieu}}, \bibinfo {author}
  {\bibfnamefont {D.}~\bibnamefont {L'H\^ote}}, \ and\ \bibinfo {author}
  {\bibfnamefont {G.}~\bibnamefont {Tarjus}},\ }\href@noop {} {\bibfield
  {journal} {\bibinfo  {journal} {Phys. Rev. E}\ }\textbf {\bibinfo {volume}
  {76}},\ \bibinfo {pages} {041510} (\bibinfo {year} {2007})}\BibitemShut
  {NoStop}%
\bibitem [{\citenamefont {Berthier}\ \emph {et~al.}(2005)\citenamefont
  {Berthier}, \citenamefont {Biroli}, \citenamefont {Bouchaud}, \citenamefont
  {Cipelletti}, \citenamefont {{El Masri}}, \citenamefont {L'H\^ote},
  \citenamefont {Ladieu},\ and\ \citenamefont {Pierno}}]{Berthier05}%
  \BibitemOpen
  \bibfield  {author} {\bibinfo {author} {\bibfnamefont {L.}~\bibnamefont
  {Berthier}}, \bibinfo {author} {\bibfnamefont {G.}~\bibnamefont {Biroli}},
  \bibinfo {author} {\bibfnamefont {J.-P.}\ \bibnamefont {Bouchaud}}, \bibinfo
  {author} {\bibfnamefont {L.}~\bibnamefont {Cipelletti}}, \bibinfo {author}
  {\bibfnamefont {D.}~\bibnamefont {{El Masri}}}, \bibinfo {author}
  {\bibfnamefont {D.}~\bibnamefont {L'H\^ote}}, \bibinfo {author}
  {\bibfnamefont {F.}~\bibnamefont {Ladieu}}, \ and\ \bibinfo {author}
  {\bibfnamefont {M.}~\bibnamefont {Pierno}},\ }\href@noop {} {\bibfield
  {journal} {\bibinfo  {journal} {Science}\ }\textbf {\bibinfo {volume}
  {310}},\ \bibinfo {pages} {1797} (\bibinfo {year} {2005})}\BibitemShut
  {NoStop}%
\bibitem [{\citenamefont {Crauste-Thibierge}\ \emph {et~al.}(2010)\citenamefont
  {Crauste-Thibierge}, \citenamefont {Brun}, \citenamefont {Ladieu},
  \citenamefont {L'H\^ote}, \citenamefont {Biroli},\ and\ \citenamefont
  {Bouchaud}}]{Thibierge10}%
  \BibitemOpen
  \bibfield  {author} {\bibinfo {author} {\bibfnamefont {C.}~\bibnamefont
  {Crauste-Thibierge}}, \bibinfo {author} {\bibfnamefont {C.}~\bibnamefont
  {Brun}}, \bibinfo {author} {\bibfnamefont {F.}~\bibnamefont {Ladieu}},
  \bibinfo {author} {\bibfnamefont {D.}~\bibnamefont {L'H\^ote}}, \bibinfo
  {author} {\bibfnamefont {G.}~\bibnamefont {Biroli}}, \ and\ \bibinfo {author}
  {\bibfnamefont {J.-P.}\ \bibnamefont {Bouchaud}},\ }\href@noop {} {\bibfield
  {journal} {\bibinfo  {journal} {Phys. Rev. Lett.}\ }\textbf {\bibinfo
  {volume} {104}},\ \bibinfo {pages} {165703} (\bibinfo {year}
  {2010})}\BibitemShut {NoStop}%
\bibitem [{\citenamefont {Bauer}\ \emph {et~al.}(2013)\citenamefont {Bauer},
  \citenamefont {Lunkenheimer},\ and\ \citenamefont {Loidl}}]{Bauer13}%
  \BibitemOpen
  \bibfield  {author} {\bibinfo {author} {\bibfnamefont {{\relax
  Th}.}~\bibnamefont {Bauer}}, \bibinfo {author} {\bibfnamefont
  {P.}~\bibnamefont {Lunkenheimer}}, \ and\ \bibinfo {author} {\bibfnamefont
  {A.}~\bibnamefont {Loidl}},\ }\href@noop {} {\bibfield  {journal} {\bibinfo
  {journal} {Phys. Rev. Lett.}\ }\textbf {\bibinfo {volume} {111}},\ \bibinfo
  {pages} {225702} (\bibinfo {year} {2013})}\BibitemShut {NoStop}%
\end{thebibliography}%

\end{document}